\documentclass{elsart}

 \usepackage{graphicx}

\usepackage{amssymb}

\begin{document}

\begin{frontmatter}



\title { First-order transition of a compartmentalized surface model for fluid membranes}


\author[label1]{Hiroshi Koibuchi}
\ead{koibuchi@mech.ibaraki-ct.ac.jp}

\address[label1]{Department of Mechanical and Systems Engineering, Ibaraki National College of Technology, 
Nakane 866, Hitachinaka,  Ibaraki 312-8508, Japan}

\begin{abstract}
Dynamically triangulated surface model is found to undergo a first-order crumpling transition between the smooth phase and a crumpled phase. Triangulated spheres are divided into compartmentalized domains, whose boundary bonds remain unflipped in the Monte Carlo simulations. The vertices can freely diffuse only inside the compartments, and they are prohibited from the free diffusion over the surface due to the domain boundaries. The violation of reparametrization invariance of the Hamiltonian is a reason for the existence of the long-range order in the fluid surface model.
\end{abstract}

\begin{keyword}
Phase Transition \sep Extrinsic Curvature \sep Elastic Membranes 
\PACS  64.60.-i \sep 68.60.-p \sep 87.16.Dg
\end{keyword}
\end{frontmatter}

\section{Introduction}\label{intro}
The crumpling transition of membranes is an interesting topic in the softmatter physics as well as in the biological physics \cite{NELSON-SMMS2004,Gompper-Schick-PTC-1994,Bowick-PREP2001}. A well-known model for such transition is the surface model of Helfrich, Polyakov, and Kleinert \cite{HELFRICH-1973,POLYAKOV-NPB1986,KLEINERT-PLB1986}. A considerable number of theoretical and numerical studies have been devoted to reveal the phase structure of the model \cite{Peliti-Leibler-PRL1985,DavidGuitter-EPL1988,PKN-PRL1988,KANTOR-NELSON-PRA1987,KD-PRE2002,KOIB-PRE-2005,KOIB-NPB-2005,Baum-Ho-PRA1990,CATTERALL-NPBSUP1991,AMBJORN-NPB1993,KOIB-EPJB-2005}. Recently, it was shown by Monte Carlo (MC) simulations that the model undergoes a first-order transition on spherical and fixed connectivity surfaces \cite{KD-PRE2002,KOIB-PRE-2005}, and the transition is universal \cite{KOIB-NPB-2005}. The vertices can move only locally on the surface because of the fixed connectivity nature in those surface models.

However, the crumpling transition is not yet clearly understood in biological membranes. If we consider the possibility of the transition in the cell membranes, we should take account of the fluid nature such as the lateral diffusion of lipids. 

Conventionally, the free diffusion of lipids has been realized by the dynamical triangulation technique in the surface model\cite{Baum-Ho-PRA1990,CATTERALL-NPBSUP1991,AMBJORN-NPB1993,KOIB-EPJB-2005}. The diffusion of lipids corresponds to the reparametrization invariance of the model, because the diffusion has no cost in energy on fluid surfaces. 

In the cell membranes, however, the free diffusion of lipids is suffered from heterogeneous structures. Lipid molecules of the cell membranes are known to undergo the so-called hop diffusion over the surface, which was recently found experimentally \cite{Kusumi-BioJ-2004}. The free diffusion of lipids is prohibited due to the cytoskeleton. The diffusion rate is, therefore, 10-100 times lower than that of artificial membranes, which is usually homogeneous and has no such domain structure. Moreover, some artificial membranes are considered to have skeletons, because they are partly polymerized \cite{CNE-PRL-2006}. 

Motivated by this fact observed in the cell membranes, we study in this Letter a dynamically triangulated surface model with compartmentalized domains whose boundaries are consisting of triangle edges (or bonds) that are not to be flipped. The diffusion is constrained so that vertices can diffuse only inside each domain, and hence the vertices never jump across from one domain to the other domains. Nevertheless, we consider that such constrained lateral diffusion can simulate the hop diffusion in the cell membranes as a first approximation.  

\section{Model}\label{model}
By dividing every edge of the icosahedron into $L$-pieces of the uniform length, we have a triangulated surface of size $N\!=\!10L^2\!+\!2$ (= the total number of vertices). The starting configurations are thus characterized by $N_5\!=\!12$ and $N_6\!=\!N\!-\!12$, where $N_q$ is the total number of vertices with the co-ordination number $q$. 

The compartmentalized structure is built on the surface by keeping the boundary bonds unflipped in the MC simulations with dynamical triangulation. The boundary of the compartment is constructed from a sequence of bonds that remain unflipped. The total number $N_C$ of the compartments depends on the surface size $N$. We fix $n$ the total number of vertices inside a compartment to the following three values:
\begin{equation}
\label{number-inside}
n=21,\;36,\;66\;\; (\# \;{\rm of\; vertices\; in \;a \;compartment}).
\end{equation}
As a consequence, $N_C$ is increased with the increasing $N$. The reason why we fix $n$ is that the size of compartment is considered to be finite, and then it is expected that total number of lipids in the compartment also remains finite in the cell membranes. We must emphasize that the finiteness of $n$ is physically meaningful rather than the value of $n$, because we do not always have one to one correspondence between the vertices and the lipid molecules.  

\begin{figure}[htb]
\unitlength 0.1in
\begin{picture}( 0,0)(  0,0)
\put(6,-1.5){\makebox(0,0){(a) $(N,n)\!=\!(2562,21)$ }}%
\put(28,-1.5){\makebox(0,0){(b) $(N,n)\!=\!(15212,66)$ }}%
\end{picture}%
\vspace{0.5cm}
\centering
\includegraphics[width=10.5cm]{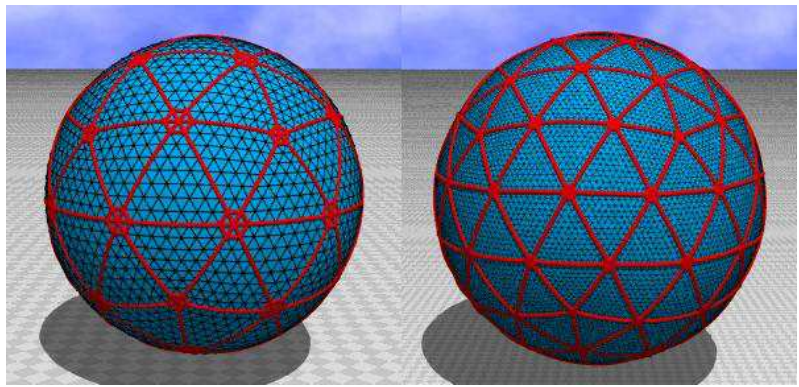}
\caption{Starting configuration of surfaces of (a) $(N,n)\!=\!(2562,21)$ and  (b) $(N,n)\!=\!(15212,66)$, where $n$ is the total number of vertices inside one compartment. Thick lines denote the compartment boundary consisting of the bonds not to be flipped.} 
\label{fig-1}
\end{figure}

Figures \ref{fig-1}(a),(b) show surfaces of $(N,n)\!=\!(2562,21)$ and $(N,n)\!=\!(15212,66)$  for the starting configurations of MC simulations. Thick lines denote the compartment boundary consisting of the bonds that are not to be flipped. Vertices on the boundary of compartments can locally fluctuate, and they are prohibited from the diffusion. The other vertices freely diffuse only inside the compartment.

We note that the tethered surface model is obtained from the compartmentalized model in the limit $n\!\to\!1$, where the vertices are prohibited from the free diffusion. On the contrary, we obtain the fluid surface model in the limit of $n\!\to\!N$, where all the vertices freely diffuse over the surface.

The surface model is defined by the partition function 
\begin{eqnarray} 
\label{Part-Func}
 Z = \sum_{ T} ^\prime\int^\prime \prod _{i=1}^{N} d X_i \exp\left[-S(X,{ T})\right],\\  
 S(X,{ T})=S_1 + b S_2, \nonumber
\end{eqnarray} 
where $b$ is the bending rigidity, $\int^\prime$ denotes that the center of the surface is fixed in the integration. $S(X,{ T})$ denotes that the Hamiltonian $S$ depends on the position variables $X$ of the vertices and the triangulation ${ T}$. $\sum_{ T}^\prime$ denotes the sum over all possible triangulations ${ T}$, which keeps the compartments unflipped. The Gaussian term $S_1$ and the bending energy term $S_2$ are defined by
\begin{equation}
\label{Disc-Eneg} 
S_1=\sum_{(ij)} \left(X_i-X_j\right)^2,\quad S_2=\sum_{(ij)} (1-{\bf n_i} \cdot {\bf n_j}),
\end{equation} 
where $\sum_{(ij)}$ in $S_1$ is the sum over bonds $(ij)$ connecting the vertices $i$ and $j$, and $\sum_{(ij)}$ in $S_2$ is also the sum over bonds $(ij)$, which are edges of the triangles $i$ and $j$. ${\bf n_i}$ in Eq. (\ref{Disc-Eneg}) is the unit normal vector of the triangle $i$.

\section{Monte Carlo technique}\label{MC-Techniques}
The vertices $X$ are shifted so that $X^\prime \!=\! X\!+\!\delta X$, where $\delta X$ is randomly chosen in a small sphere. The new position $X^\prime$ is accepted with the probability ${\rm Min}[1,\exp(-\Delta S)]$, where $\Delta S\!=\! S({\rm new})\!-\!S({\rm old})$. 

The summation over ${ T}$ in $Z$ of Eq.(\ref{Part-Func}) is performed by the standard technique for bond flip. The bonds are labeled with sequential numbers. The total number of bonds $N_B$ is given by $N_B\!=\!3N\!-\!6$, which includes the bonds making the boundary of compartments.  Firstly, the odd-numbered bonds are sequentially chosen to be flipped, and secondly the remaining even-numbered bonds are chosen. The flip is accepted with the probability ${\rm Min} [1, \exp(-\Delta S)]$. In this procedure, the compartment boundary remains unflipped. $N$-updates for $X$ and $N_B/2$-updates for ${ T}$ are consecutively performed and make one MCS (Monte Carlo Sweep). The radius of the small sphere for $\delta X$ is chosen so that the rate of acceptance for $X$ is about $50\%$. We introduce the lower bound $1\times 10^{-8}$ for the area of triangles. No lower bound is imposed on the bond length. We use surfaces of size $N\!=\!2562$, $N\!=\!5762$, and $N\!=\!10242$ for $n\!=\!21$,  and $N\!=\!1002$, $N\!=\!4002$, and $N\!=\!9002$ for $n\!=\!36$, and $N\!=\!1692$, $N\!=\!6762$, and $N\!=\!15212$ for $n\!=\!66$. 

\section{Results}\label{Results}
\begin{figure}[htb]
\centering
\includegraphics[width=10.5cm]{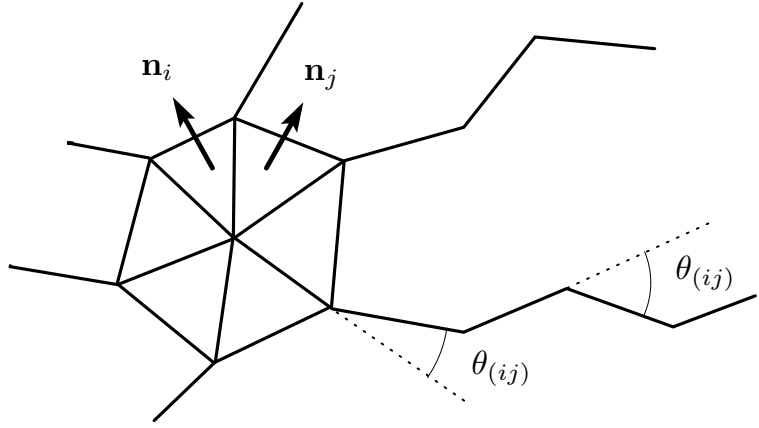}
\caption{The mean square size $X^2$ vs. $b$ obtained on the surfaces of (a) $n\!=\!21$, (b) $n\!=\!36$, and (c) $n\!=\!66$. Solid lines were obtained by multihistogram reweighting technique.} 
\label{fig-2}
\end{figure}

The mean square size $X^2$ is defined by
\begin{equation}
\label{X2}
X^2={1\over N} \sum_i \left(X_i-\bar X\right)^2, \quad \bar X={1\over N} \sum_i X_i,
\end{equation}
where $\bar X$ is the center of the surface. Figures \ref{fig-2}(a),(b), and (c) show $X^2$ obtained on the surface of $n\!=\!21$, $n\!=\!36$, and $n\!=\!66$, respectively. Solid lines connecting data were obtained by multihistogram reweighting technique \cite{Janke-histogram-2002}. Discontinuous change of $X^2$ can be seen in each case of $n$ when the size $N$ increases, and indicates a first-order transition. We also find that the transition point $b_n$ moves left on the $b$ axis as $n$ decreases. It is expected in the limit of $n\!\to\! 1$ that $b_n$ reduces to the value, which corresponds to the transition point of the tethered surface model \cite{KOIB-PRE-2005}. On the contrary, $b_n$ moves right on the $b$-axis as $n$ increases. We expect that $b_n$ disappears in the limit of $n\!\to\!N$ at sufficiently large $N$, which will be confirmed later in this Letter.

\begin{figure}[htb]
\centering
\includegraphics[width=10.5cm]{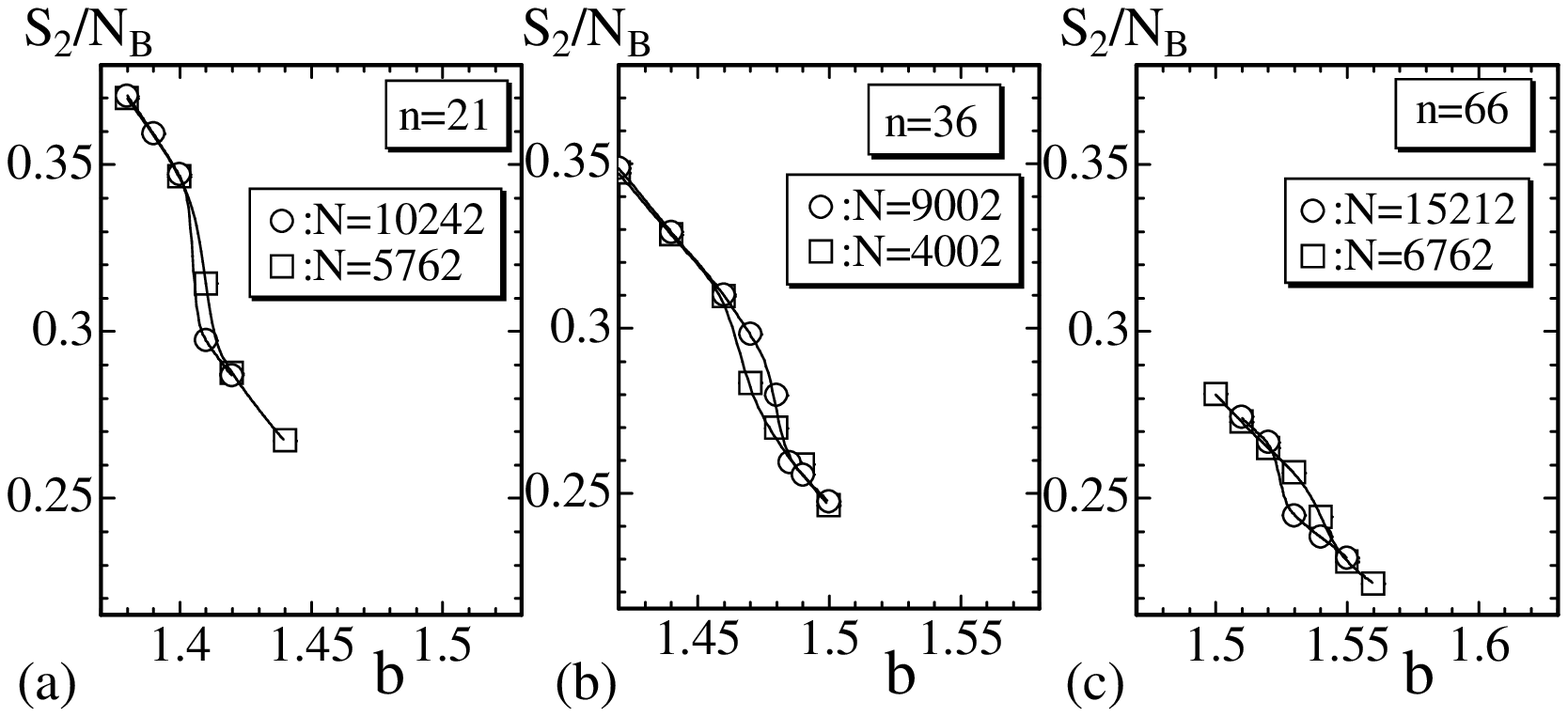}
\caption{The bending energy $S_2/N_B$ vs. $b$ obtained on the surface of (a) $n\!=\!21$, (b) $n\!=\!36$, and (c) $n\!=\!66$. $N_B(\!=\!3N\!-\!6)$ is the total number of bonds.  } 
\label{fig-3}
\end{figure}
The bending energy $S_2/N_B$ defined in Eq.(\ref{Disc-Eneg}) is shown in Figs. \ref{fig-3}(a),(b), and (c), which correspond to $n\!=\!21$, $n\!=\!36$, and $n\!=\!66$, respectively. Discontinuous change of $S_2/N_B$ can also be seen at $b$ where $X^2$ discontinuously changes, although the discontinuity of $S_2/N_B$ is not sufficiently clear in the figures. We find also that the value of $S_2/N_B$ in the smooth phase (or in the crumpled phase) increases as $n$ decreases at the transition point. The value of $S_2/N_B$ at the transition point becomes identical to the one of the tethered surface model in the limit of $n\!\to\! 1$ \cite{KOIB-PRE-2005}. On the contrary, the discontinuous change of $S_2/N_B$ is expected to disappear at sufficiently large $n$, because the first-order transition at $n\!=\!21\sim 66$ disappear when $n\!\to \!N$ at sufficiently large $N$.

\begin{figure}[htb]
\centering
\includegraphics[width=10.0cm]{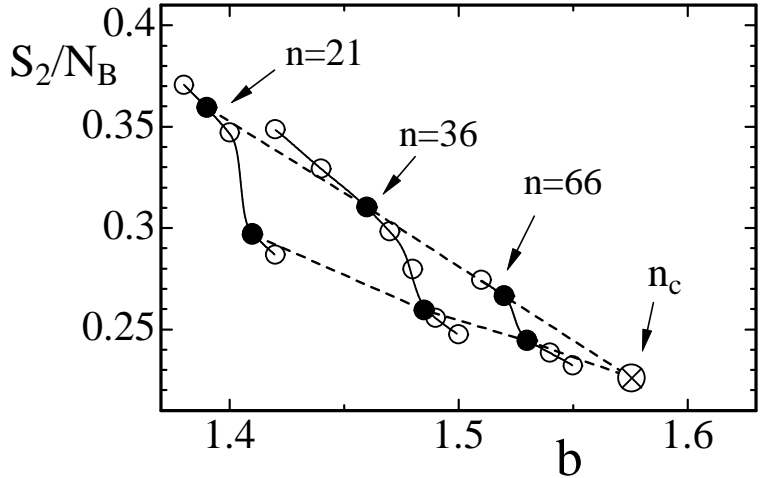}
\caption{The variation of the gap of $S_2/N_B$ against $n$, which are obtained on the $(N,n)\!=\!(10242,21)$, $(N,n)\!=\!(9002,36)$, $(N,n)\!=\!(15212,66)$ surfaces.  The solid circles ($\bullet$) and the dashed lines denote the value of  $S_2/N_B$ in the smooth (the crumpled) phase at the first-order transition point. The symbol ($\otimes $) is unknown critical value of $n$ denoted by $n_c$, where the discontinuous transition is expected to disappear. The position of $n_c$ is also unknown on the $b$-axis. } 
\label{fig-4}
\end{figure}
This implies that there exists a finite $n_c$, where the first-order transition turns to a continuous or a higher-order one. Figure \ref{fig-4} shows the variation of the gap of $S_2/N_B$ against $n$. The gap of $S_2/N_B$ at $n\!=\!21$ reduces as $n$ increases and eventually goes to zero at $n\!=\!n_c$, which will be clarified in the future. 

We note that the maximum co-ordination number $q_{\rm max}$ is as follows: $q_{\rm max}\!=\!38$ on the $(N,n)\!=\!10242$ surface at $b\!=\!1.38$, $q_{\rm max}\!=\!42$ on the $(N,n)\!=\!9002$ surface at $b\!=\!1.42$, and $q_{\rm max}\!=\!40$ on the $(N,n)\!=\!15212$ surface at $b\!=\!1.51$. These were obtained throughout the MC simulations.

\begin{figure}[htb]
\centering
\includegraphics[width=10.5cm]{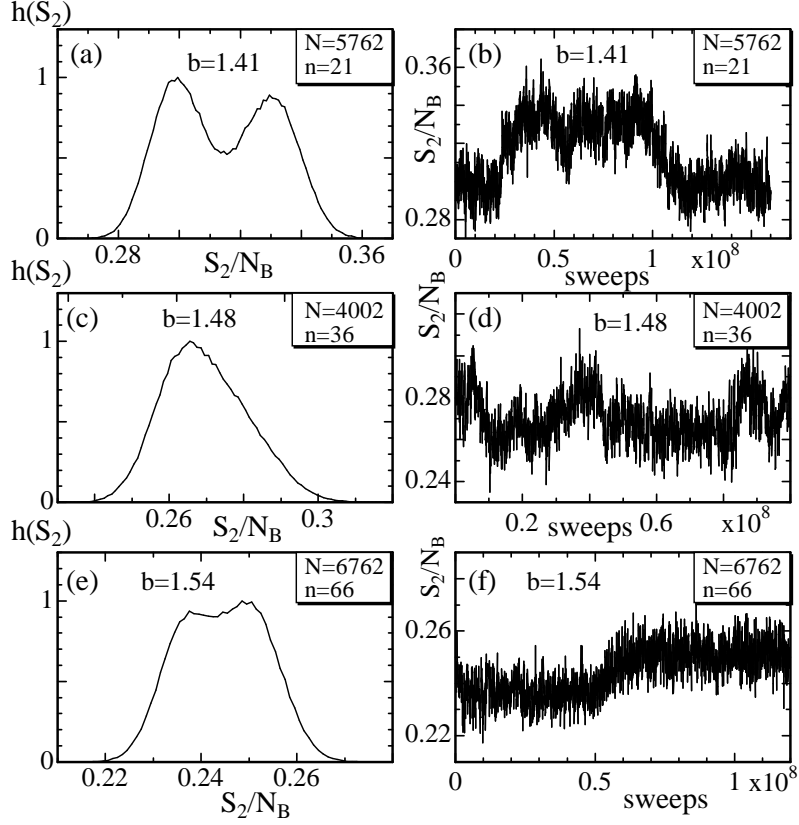}
\caption{The histogram $h(S_2)$ and the corresponding variation of $S_2/N_B$ against MCS obtained on the surfaces of (a),(b) $(N,n)\!=\!(5762,21)$,  (c),(d) $(N,n)\!=\!(4002,36)$, and (e),(f) $(N,n)\!=\!(6762,66)$.   } 
\label{fig-5}
\end{figure}
In order to show the discontinuity in $S_2/N_B$ more clearly, we plot in Figs. \ref{fig-5}(a)--\ref{fig-5}(f) the distribution (or histogram) $h(S_2)$ of $S_2/N_B$ and the corresponding variation of $S_2/N_B$. They were obtained on the surfaces of $(N,n)\!=\!(5762,21)$,  $(4002,36)$, and $(6762,66)$. The discontinuity of $S_2/N_B$ can be seen in the histogram on the $N\!\geq\!5762$ surfaces. Because of the size effect, the transition appears to be continuous on the $N\!=\!4002$ surface. The double peak at $(N,n)\!=\!(5762,21)$ is more clear than that at $(N,n)\!=\!(6762,66)$, because the gap of $S_2/N_B$ reduces as $n$ increases.   We should note that a double peak structure is very hard to see in $h(S_2)$ on the $N\!\geq\!9002$ surfaces. When the configuration is once trapped in the smooth (the crumpled) state it hardly changes to the crumpled (the smooth) state at the transition point on such large surfaces. This problem may be resolved with more sophisticated MC techniques than the current one in this Letter.

\begin{figure}[htb]
\centering
\includegraphics[width=10.5cm]{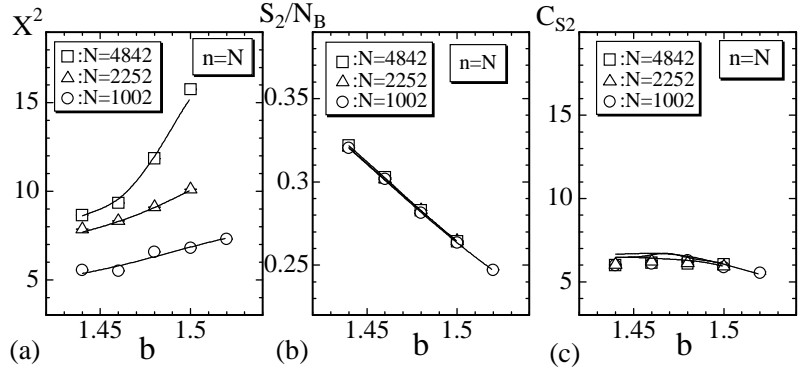}
\caption{(a) $X^2$ vs. $b$, (b) $S_2/N_B$ vs. $b$, and (c) $C_{S_2}$ vs. $b$ obtained on the fluid surface model without the compartmentalized structure. The symbol $n\!=\!N$ drawn on the figures denotes that the surfaces are those without the compartment.} 
\label{fig-6}
\end{figure}
The phase transition is expected to disappear from the fluid surface model defined on the surfaces that have no compartment. In order to show this, we performed MC simulations on the surfaces without the compartments up to the size $N\!=\!4842$. Figure \ref{fig-6}(a) shows $X^2$ against $b$. We can see no abrupt growing of $X^2$ in the figure. The bending energy $S_2/N_B$ shown in Fig. \ref{fig-6}(b), where the variation of $S_2$ against $b$ seems almost independent of $N$. The specific heat $C_{S_2}$, which is defined by 
$C_{S_2} \!=\! {b^2\over N} \langle \; \left( S_2 \!-\! \langle S_2 \rangle\right)^2\rangle$, 
is expected to reflect the phase transition. However, we can see no anomalous behavior in $C_{S_2}$ shown in Fig. \ref{fig-6}(c); there can be seen no peak in $C_{S_2}$. Thus we confirmed that the phase transition disappears from the model if $n\!\to \! N$ at sufficiently large $N$.

\begin{figure}[htb]
\unitlength 0.1in
\begin{picture}( 0,0)(  0,0)
\put(6,-1.5){\makebox(0,0){(a) $N\!=15212$, $n\!=\!66$ }}%
\put(28,-1.5){\makebox(0,0){(b) the surface section}}%
\end{picture}%
\vspace{0.5cm}
\centering
\includegraphics[width=10.5cm]{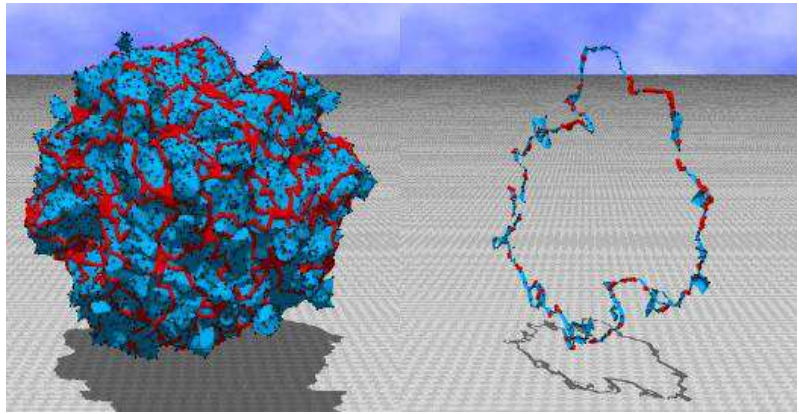}
\caption{Snapshot of the surface of $(N,n)\!=\!(15212,66)$ obtained in the smooth phase at (a) $b\!=\!1.53$, which is close to the transition point, and (b) the corresponding surface section. } 
\label{fig-7}
\end{figure}
Finally, we show in Fig. \ref{fig-7}(a) a snapshots of the $(N,n)\!=\!(15212,66)$ surface, which is obtained in the smooth phase close to the transition point.  Figure \ref{fig-7}(b) shows the surface section. Thus we confirmed that the smooth phase can be seen close to the transition point at finite $b$ in contrast to the case $n\!\to\!N$, where it can be seen only at $b\!\to\!\infty$.

\section{Summary and Conclusion}
To the summary, we have studied a compartmentalized surface model for the cell membranes and found that the model undergoes a first-order transition between the smooth phase and the crumpled phase. The model is classified as a fluid surface model, although the long-range order and the phase transition can be seen at finite $b$. The compartmentalized structure is an origin of a broken symmetry for the reparametrizaition invariance, which is considered to be a reason for the existence of the phase transition. Consequently, the result is not in contradiction with the standard argument for the non-existence of long-range order in fluid membranes.  Moreover, the critical point of the transition is strongly expected at finite $n_c(>66)$. The number $n_c$ and the corresponding critical bending rigidity $b_c$ remained to be clarified. However, it is possible that the compartmentalized fluid surface model undergoes a continuous transition, because the strength of the transition is expected to be continuously softened with the increasing $n$. We observed that the phase transition remains first-order at $n\!=\!66$ and disappear at $n\!\to\!N$ in the limit of $N\!\to\!\infty$. Although the demonstration of the continuous transition is computationally a hard task, it will be performed and the results including those in this Letter will be reported elsewhere.

This work is supported in part by a Grant-in-Aid for Scientific Research, No. 15560160.  



\end{document}